\documentstyle[preprint,aps]{revtex}

\begin{document}
\draft
\title{\bf Optical scalars in  spherical spacetimes}
\author{E. Malec$^{+*}$ and N. \'O Murchadha$^+$}
\address{$^+$Physics Department, University College, Cork,
Ireland}
\address{$^*$Uniwersytet Jagiello\'nski, Instytut Fizyki,
30-059 Krak\'ow, Poland}

\maketitle
\begin{abstract}
Consider a spherically symmetric
spacelike slice through a spherically symmetric spacetime. One can
derive a universal bound for the optical scalars on any such slice. The only
requirement is  that the matter sources satisfy the dominant energy condition
and that the slice be asymptotically flat and regular at the origin. This
bound  can
be used to derive new conditions for the formation of apparent horizons.
 The bounds hold even when the matter has a distribution on
a shell or blows up at the origin so as to give a conical singularity.
   \end{abstract}    \date{\today}
\narrowtext
Relativists, especially those who are numerically inclined \cite{9}, have
long known
that regular spacelike slices often wrap around singularities
rather than approaching them. In this letter we derive a new and
remarkable relation which gives a bound on  the optical scalars and
which shows how slices which are
asymptotically flat may be prevented from coming close to
singularities.

Consider a spacelike slice through spacetime.
The geometry of this slice cannot be chosen at will; it must
satisfy the constraint equations. In the spherically
symmetric case these constraints can be written as equations for the
optical scalars. These equations, combined with the requirement of regularity
at the origin and at infinity, force the optical scalars to remain bounded
over the entire slice. The optical scalars are four-dimensional objects which
we expect to become unboundedly large as one approaches a singularity. Thus
regular spacelike slices are excluded from regions near singularities. This
bound on the optical scalars also has a more immediate use. Over the years, we
have been interested in developing criteria to determine when and if apparent
horizons form \cite{1,2}. In spherically symmetric systems the existence of
an apparent horizon implies the existence of a black hole \cite{2a,2b}. These
bounds on the optical scalars allow us sharpen significantly our condition for
the formation of apparent horizons.

  We define a spherically symmetric spacetime as one having the
metric
\begin{equation}
ds^2 = -\alpha^2(r,t) dt^2 + a(r,t)dr^2 + b(r,t)r^2 d\Omega^2
\label{1}
\end{equation}
where $0\le \phi\le 2\pi $ and $0\le \theta <\pi $ are the standard
angle variables such
that $d\Omega^2 = d\theta^2 + \sin^2\theta d\phi^2$.

The initial data for the Einstein equations are prescribed by giving
the spatial
geometry at $t=0$, i.e., by specifying the functions $a(r,0)$ and $b(r,0)$, and
by  giving the extrinsic curvature (again at $t=0$)

\begin{equation}
K_r^r={\partial_ta\over 2a \alpha },~~K_{\theta }^{\theta }=
K_{\phi }^{\phi }={\partial_tR\over \alpha R},~~tr K=
{\partial_t(\sqrt{a}b)\over  \sqrt{a}b \alpha }
\label{2}
\end{equation}
where the areal (Schwarzschild) radius $R$ is defined as

\begin{equation}
R=r\sqrt{b}
. \label{3}
\end{equation}
It is useful to define the  mean curvature of a centered sphere in
the initial hypersuface by
\begin{equation}
p= {2\partial_rR\over \sqrt{a} R}.
\label{4}
\end{equation}

In a general spacetime the behaviour of a pencil of lightrays is described by
specifying a number of functions which describe the expansion and shear of the
rays. In a spherically symmetric spacetime, however, we need specify only two.
These objects can be expressed in terms of the initial data on any spacelike
slice so they are simultaneously three-dimensional and four-dimensional
scalars.  These optical scalars are the divergence of future directed light
rays

\begin{equation}
\theta ={2\over R}{d\over \alpha dt}_{out}R=p-K_r^r+trK,
\label{5}
\end{equation}
and   the divergence of  past directed light rays

\begin{equation}
\theta ' ={-2\over R}{d\over \alpha dt}_{in}R=p+K_r^r-trK,
\label{6}
\end{equation}
where ${d\over \alpha dt}_{in}={\partial_t\over \alpha }-{\partial_r\over
 \sqrt{a}}$
and ${d\over \alpha dt}_{out}={\partial_t\over \alpha }+{\partial_r\over
\sqrt{a}}$  are  the full derivatives in the direction orthogonal to
the centered sphere of ingoing
and outgoing photons respectively. In flat
space-time both quantities are positive and equal to $2/R$, where $R$ is the
radius of a sphere; hence each of the products $R\theta $ and
 $R\theta'$ equals 2.

The initial data must satisfy the
constraints. These constraints, expressed in terms of $\theta $ and
$\theta '$, can  be written as
\begin{eqnarray}
\partial_l(\theta R)=&&-8\pi R(\rho - {j_r\over \sqrt{a}}) -{1\over 4R}
\Bigl( \theta^2R^2-4 -4\theta tr K R^2\nonumber\\
&& +\theta R (\theta R-\theta 'R)\Bigr),
\label{7}
\end{eqnarray}

\begin{eqnarray}
\partial_l(\theta 'R)=&&-8\pi R(\rho + {j_r\over \sqrt{a}}) - {1\over 4R}
\Bigl( \theta '^2R^2-4 +4\theta 'tr K R^2\nonumber\\
&& +\theta 'R (\theta 'R-\theta R)
\Bigr),
\label{8}
\end{eqnarray}
where $l$ is the proper distance from the center, i.e., $dl =\sqrt{a}dr$.
$\rho$ and $j_r$ are the energy density and the current density of the sources
that generate the gravitational field. Note that $j_r/\sqrt{a}$ equals $j.n$
where $n$ is the unit normal in the radial direction. We will assume that the
sources satisfy the dominant energy condition,  $\rho \ge  |j|  $.
If the origin is regular, local flatness forces  both optical scalars to
satisfy the conditions $\lim_{R\rightarrow 0} \theta R = \lim_{R \rightarrow 0}
\theta ' R=2$. Asymptotic flatness also gives $\lim_{R\rightarrow
\infty} \theta R = \lim_{R \rightarrow \infty} \theta ' R=2$.

The primary result of this calculation is a proof that if $\theta R$, $\theta'
R$ are
bounded at the origin and at infinity they are bounded on the entire
hypersurface.
Define $B=4 \sup_{0\le r \le \infty } (|R trK|)$.
  We prove

{\bf Lemma 1.} Given spherical initial data that are regular at the
origin and at infinity with sources that satisfy the dominant energy
condition, both optical scalars are bounded on the entire hypersurface
\begin{equation}
-2-B\le \theta R,~ \theta 'R \le 2+B.
\label{9}
\end{equation}
{\bf Proof:} Let us assume that $\theta R \ge 2 + B$ and $\theta R \ge \theta'
R$.
Consider the non-source part of eqn.(\ref{7}), i.e., $( \theta^2R^2 -  4
- 4\theta trK R^2 +\theta R (\theta R-\theta 'R)$. Since  $\theta R \ge 2 + B$,
the first three terms are nonnegative while
$\theta R \ge \theta' R$ means that the last term is nonnegative. Therefore
eqn.(\ref{7}) implies that $\partial_l(\theta R) \le 0$. Also, if
 $\theta' R \ge 2 + B$ and $\theta' R \ge \theta R$, a similar analysis of
eqn.(\ref{8}) gives  $\partial_l(\theta' R) \le 0$. Therefore, if
$\max(\theta R, \theta' R) \ge 2 + B$ then $\partial_l[\max(\theta R, \theta'
R)] \le 0$. Since the maximum starts at 2, and the derivative at $2 + B$ is
negative, it cannot rise above $2 + B$. Hence $2 + B$ is an upper bound.

The argument for the lower bound works in exactly the same way.
Let us assume that $\theta R \le -2 - B$ and that $\theta R \le
\theta' R$.  Again
eqn.(\ref{7}) means that $\partial_l(\theta R) \le 0$. Hence if
$\min(\theta R, \theta' R) \le -2 - B$ then $\partial_l[\min(\theta R, \theta'
R)] \le 0$. This means that one or the other of ($\theta R, \theta' R$) is
driven
more and more negative. However, asymptotic flatness demands that both rise up
to +2 at infinity. Contradiction!

Let us stress that while $B$ is a three-scalar which depends on the
particular spacelike slice, $\theta, \theta'$ and $R$ are
four-dimensional scalars, properties of the spacetime, which are
independent of the choice of foliation. Thus eqn.(\ref{9}) places
restrictions on the kind of regular spacelike slice that may enter
particular regions of spacetime.

 There are only two allowed topologies
for globally regular, asymptotically flat, spherically symmetric,
spacelike three-manifolds. They can either have $R^3$ topology with a
regular center and one asymptotic end or $R \times S^2$ topology with two
asymptotic ends, as in the Schwarzschild geometry. Lemma 1 holds in
both cases.

Lemma 1 has a number of interesting consequences.
Let us assume, for a moment, that the trace of the extrinsic
curvature vanishes, i.e., that the initial data define a maximal slice.
This means that $B \equiv 0$ and Lemma 1 implies that $|\theta R|,
|\theta' R| \leq 2$. A surface on which $\theta < 0$ is called a
trapped surface; such surfaces play a key role in the singularity theorems
of general relativity.
 Eqn.(\ref{7}) can be used to derive

\begin{equation}
\partial_l(\theta R^2)=-8\pi R^2(\rho - {j_r\over \sqrt{a}}) + 1
  +{1\over 4} \theta R (2\theta 'R -\theta R).
\label{10}
\end{equation}

 Let $L(S)$ be the geodesic (proper) radius of a sphere $S$; $R(S)$ its areal
radius;
 $M(S)=\int_{V(S)}\rho dV$ the total mass inside $S$; and
$P(S)=\int_{V(S)}{j_r\over
 \sqrt{a}}dV$ be the total radial momentum. Integrating  (\ref{10}) noting
that $4\pi R^2 dl = 4\pi\sqrt{a} R^2 dr$ is the proper volume,  we get

\begin{eqnarray}
 (\theta R^2)(S)
 = &&-2(M-P)(S) + L(S)\nonumber\\
&&+ {1\over 4}\int_0^{L(S)}\theta R (2\theta 'R -\theta R)dl.
\label{11}
\end{eqnarray}
We can see that
${1\over 4}\int_0^Ldl\theta R (2\theta 'R -\theta R)\le
{1\over 4}\int_0^Ldl(\theta 'R  )^2\le L$, where the first inequality
comes from the trivial estimate $2ab -a^2 \le b^2$ and the second from Lemma
1.  Therefore

\begin{equation}
 (\theta R^2)(S) \le -2(M-P)(S) +2L(S)
\end{equation}
for any surface $S$. In particular, if $M-P\ge L$ at any given sphere $S$ then
$\theta (S)$ must be negative. Thus we have proven:

{\bf Theorem 1.} Under conditions of Lemma 1, assuming $tr K \equiv 0$,
 if the difference between the
total rest mass $M(S)$  and the radial momentum $P(S)$ exceeds the proper
radius $L(S)$ of a sphere $S$,
$M(S)-P(S) > L(S)$,
then $S$ is trapped.

This theorem improves our earlier result\cite{1}, in which
we got a similar result  but with  $L$ replaced by
 ${7\over 6}L$ and the
weaker conclusion that there exists a trapped surface inside $S$. The
difference is due to the fact that  we now impose the somewhat stronger
condition that $\rho - |j| \ge 0$, whereas in \cite{1}
we used $\rho +{3\over 32\pi}(K_r^r)^2\ge 0$.    Since  the new conditions in
Theorem 1 eliminate tachyons this is a real difference.
The constant ${7\over 6}$ also appears
in our criteria for the formation of cosmological black holes \cite{2};
we  believe that these can also be improved to 1.

The  meaning  of Theorem 1 is  transparent. Radially ingoing matter
 $j_r\le 0$
helps form apparent horizons. The presence of  outgoing matter, i.e .,
when $P(S)$  becomes positive, has to be compensated  for by a  greater matter
density. In the extremal case of radially outgoing photons, when $M(S)=P(S)$,
apparent horizons cannot form. This follows from our Theorem 2  below.

Theorem 1 is sharp in the sense that there exists an initial value
configuration when the inequality saturates. This is a 3-geometry
created by a shell of moving matter; the explicit calculation will be done
elsewhere.  The case in which $P=0$ was discussed in \cite{1} and the
corresponding criterion (with the same constant 1, as above) was shown
to be the best possible.

It is interesting that we obtain an exact  criterion with   the
constant 1; this suggests that Theorem 1 is part of a
more complex true statement that can be formulated for general nonspherical
spacetimes. It suggests also that $M(S)$ is a sensible measure of the energy
of a gravitational  system that  might appear as a part of a
quasilocal energy measure in nonspherical systems.

We also obtain  a necessary condition for the formation of apparent
horizons. In  \cite{3} we found a criterion based on asymptotic
data outside a collapsing system. \cite{1}  states  that  $M(S)>{L\over 2}$
must be satisfied if $S$ is trapped in the case of moment of time symmetry
data.  The same holds true if the matter is moving under some stringent
conditions on the sign of the momentum density \cite{4}. Here we will derive a
different (and not  particularly interesting, although exact)
estimate. The most
important assumption we make is that $\theta'$ is everywhere positive on the
initial hypersurface. Just as $\theta \le 0$ guarantees a singularity to the
future, $\theta' \le 0$
guarantees a singularity to the past. Therefore, data which arises from a
regular past must have positive $\theta'$.

{\bf Theorem 2.} Assume a regular maximal slice on which the sources satisfy
the dominant energy condition.  Let  $S$  be the innermost trapped surface and
let
$ (R\theta ') > \epsilon >0$  inside $S$. Then
 $$M(S)-P(S) \ge {\epsilon \over 2}L.$$
{\bf Proof.} As before, we consider
(\ref{11}), which reads
\begin{equation}
 \theta R^2 =-2(M-P) + L+ {1\over 4}\int_0^Ldl\theta R (2\theta 'R
-\theta R).\label{13}
\end{equation}
Inside $S$,   $R\theta $ is positive. We seek a lower bound on
the last term on the right hand side of (\ref{13}). Let $t=R\theta ,
u=R\theta '$; from Lemma 1 we know $\mid t \mid , \mid u \mid \le 2$,
so our task
consists in estimating   $2tu-t^2$ for $0\le t\le 2, \epsilon \le u \le 2$. We
know that $2tu-t^2 \ge F(t) = 2t\epsilon - t^2$. The only extremum of
 $F(t)$ is a
maximum at $t = \epsilon$.  The minimum must occur at the endpoints and it is
easy to show that $2tu -t^2 \ge F(t) \ge  4\epsilon  - 4$. Inserting
this  into
(\ref{13}) yields

\begin{eqnarray}
 \theta (S) R^2 &\ge& -2(M-P)(S) + L(S)+
 {1\over 4}\int_0^{L(S)}dl(4\epsilon -4)\nonumber\\
&=& -2(M-P)(S)+\epsilon L,
\label{14}
\end{eqnarray}
that is, since $\theta (S)=0$, $$ M(S)-P(S)\ge {\epsilon L \over 2}.$$
Hence Theorem 2 is proven.

The inequality of Theorem 2  becomes an equality in the case of a spherical
shell.
The   geometry inside the shell is flat and $\theta 'R=2$. The necessary
condition that the shell be trapped is that $M-P>L$. In
\cite{1} we proved this in the special case when $P=0$.

It is clear that the analysis performed here can include cases where
 the sources
are distributions rather than classical functions; in particular, we have no
difficulty with shells of matter. All we get on crossing the shell is a
downward step in $\theta$ and $\theta'$. More interestingly, we can extend
the analysis to include weak singularities at the origin.

Let us begin by considering a conical singularity\cite{5}. Consider a metric of
the form \begin{equation}
dS^2 = dr^2  + a^2r^2 d\Omega^2.
\end{equation}
The scalar curvature of this metric is $^{(3)}R = 2(1 - a^2)/r^2$. A moment of
time symmetry data set is one for which $j^i$ and $K^{ij} \equiv 0$. For such
data sets the constraints reduce to  $^{(3)}R = 16\pi\rho$. For the above
metric we get $\rho = (1 - a^2)/8\pi r^2$. The dominant energy condition
reduces
 to
the positivity of $\rho$, which implies $a^2 \leq 1$. For this metric we can
also compute the mean curvature $p$, which in this case equals both $\theta$
and $\theta'$, to get $p = 2/r = 2a/R$. Hence we get $|pR| \leq 2$.
However, the argument of Lemma 1 only requires that $\theta R, \theta'
R$ be bounded at the origin. Therefore we have shown that Lemma 1 holds
for moment of time symmetry data with a conical singularity at the
origin. The conical singularity in question is determined by the deficit
of the solid angle $4\pi (1- a^2)$. We will show that  a similar  result holds
true for general nonmaximal data.

Let us consider initial data such that $tr K$ is finite while  $R\theta
\rightarrow X$ and $R\theta' \rightarrow Y$ as $R \rightarrow 0$. Let us also
assume that $\partial_l(R\theta)$ and $\partial_l(R\theta')$ are finite at $R =
0$. There are terms on the right hand side of eqns.(7) and (8) which seem to
diverge like 1/R. The source term will have the same sort of 1/R divergence if
$8\pi R^2\rho \rightarrow \alpha$ and $8\pi R^2 j_r/\sqrt{a}
\rightarrow \beta$,
just as in the case of the conical singularity. The coefficient of this $1/R$
term must vanish. This gives us a pair of equations, one from (7) and one from
(8) \begin{equation}
\alpha - \beta + {1 \over 2}X^2 - {1 \over 4}XY - 1 = 0;
\end{equation}
\begin{equation}
\alpha + \beta + {1 \over 2}Y^2 - {1 \over 4}XY - 1 = 0.
\end{equation}

By adding these equations we get
\begin{equation}
4\alpha = 4 - X^2 - Y^2 + XY;\label{16}
\end{equation}
and by subtracting
\begin{equation}
4\beta = X^2 - Y^2.\label{17}
\end{equation}

Note that eqn.(\ref{16}) implies that $\alpha \le 1$. The weak energy
condition gives $\alpha \ge |\beta|$. Let us assume that $\beta \ge 0$.
Eqn.(\ref{17}) now gives us $X^2 \ge Y^2$ and  $Y = \pm \sqrt{X^2 - 4\beta}$.
Substituted this into eqn.(\ref{16}) to give
\begin{equation}
[3X^2 - 4(1 - \alpha + \beta)][X^2 - 4(1 - \alpha + \beta)] + 4X^2\beta^2 = 0.
\label{20}
\end{equation}

The roots of this equation, if it has any, must lie in the range
$4(1 - \alpha + \beta)/3 \le X^2 \le 4(1 - \alpha + \beta)$.
Therefore we have shown that
$2 \ge |X| \ge |Y|$. If we assume $\beta < 0$, we just reverse the roles of
$X$ and $Y$. Hence we obtain

{\bf Lemma 2.} Given $\rho \ge |j| $ and if all of $trK,
\theta R, \theta ' R, \partial_l \theta R, \partial_l \theta'R,
(8\pi \int_0^R\rho \tilde R^2d \tilde R)/R $
are finite in the limit $R=0$ then

\begin{equation}
2\ge \lim_{R \rightarrow 0}  |\theta R|, \lim_{R\rightarrow 0}|\theta 'R|,~~~
1\ge {8\pi \int_0^R\rho \tilde R^2d \tilde R \over R}.
\label{21}
\end{equation}

{}From eqns.(4), (5) and (6) it is clear that
\begin{equation}
2\partial_lR = {2\partial_rR \over \sqrt{a}} = pR = {\theta R + \theta' R \over
2}. \end{equation}
This means that the spatial part of the metric (1) can be written, at
least  in a small neighbourhood of $R=0$, as
\begin{equation}
{16\over (R\theta +R\theta ')^2}dR^2+R^2d\Omega^2.
\end{equation}
The estimate derived in lemma 2 implies that, under the stated conditions,
there  can  be at most  a conical singularity
at the origin, with solid angle deficit $4\pi (1- {(X+Y)^2\over
16})$.
Conical singularities have previously been investigated in 2+1 gravity
\cite{6}.
In the 2+1 case the conical singularity can also be described by an angle
deficit expressed in terms of the mean curvature: $2\pi (1-pR)$. However, in
the 2+1 case the geometry is locally flat but globally nontrivial and the
deficit angle is related to a total mass \cite{6}. In our case, the deficit
angle is a local phenomenon caused by a mildly singular mass distribution at
the origin, where $\rho$ diverges like $r^{-2}$.

Lemma 2 gives the desired  bound  $|\theta R|, |\theta 'R|\le 2$ at
the origin so we get a generalized version of Lemma 1:

{\bf Lemma 1$'$}.  Assume an asymptotically flat nonmaximal slice,
satisfying the
dominant energy condition, such that $4\sup_{0\le R \le \infty } |RtrK|=B$
is finite.
Let the conditions of Lemma 2 be satisfied at the origin.   Then

\begin{equation}
2+B\ge    |\theta R|,  |\theta 'R|.
\label{24}
\end{equation}
Theorems 1 and 2 hold under similar conditions.

As we have mentioned earlier, $\theta R$ and $\theta' R$ are defined
for any point in a spherically symmetric spacetime geometry,
independent of any foliation or choice of time. One consequence of
Lemma 1 is that if a point exists in a spherical spacetime for which
either $|\theta R|$ or $|\theta' R|$ is larger than 2 then we know
that a regular, maximal, asymptotically flat slice cannot pass through
this point.

Consider regular, asymptotically flat, spherically symmetric initial
data which contain an apparent horizon. Let us now evolve the spacetime
and look at the maximal Cauchy development of this data. We are guaranteed that
a
singularity will occur for
a sufficiently large value of local proper time. It seems to us that
there are only three realistic outcomes:

i) The singularity will be of the kind where $R\theta \rightarrow
-\infty$, as in the Schwarzschild singularity. Maximal slices (and any other
slicing with a regular trace of the extrinsic curvature) do not cover the full
Cauchy evolution. We get a ``collapse of the lapse''. The foliation can
continue
for infinite time as seen by asymptotic observers but ``freezes'' near the
interior.

ii) We get some sort of ``bag of gold'' forming, where $R$ goes to zero at some
finite value of the proper radius, $L$, and part of the spacetime pinches off
from the rest.

iii) A singularity appears with diverging mass density. This may be a
shell-crossing singularity, a central conical singularity as we discussed above
or some sort of `strong' central singularity. We expect that the appearance
(or otherwise) of these singularities would be determined by the matter
equation
of state.

{\bf Acknowledgement:} We would like to thank Piotr Chru\'sciel for
 helpful comments.

\end{document}